\documentclass[apj]{emulateapj}

\shorttitle{Coronal source of line asymmetries}                 
\shortauthors{Brooks \& Warren}
\begin{document}

\title{The coronal source of extreme-ultraviolet line profile asymmetries in solar active region outflows.}
\author{David H. Brooks \altaffilmark{1}}
\affil{College of Science, George Mason University, 4400 University Drive, Fairfax, VA 22030}
\altaffiltext{1}{Current address: Hinode Team, ISAS/JAXA, 3-1-1 Yoshinodai, Chuo-ku, Sagamihara, Kanagawa 252-5210, Japan}
\email{dhbrooks@ssd5.nrl.navy.mil}
\author{Harry P. Warren}
\affil{Space Science Division, Naval Research Laboratory, Washington, DC 20375}

\begin{abstract}
High resolution spectra 
from the 
Hinode EUV Imaging Spectrometer (EIS) have revealed that coronal 
spectral line profiles are sometimes asymmetric, with a faint enhancement in the 
blue wing on the order of 100\,km s$^{-1}$. These asymmetries could be important since
they may be subtle, yet diagnostically useful signatures of coronal heating or solar wind acceleration 
processes. It has also been suggested that they are signatures of 
chromospheric jets supplying mass and energy to the corona. Until now, however,
there have been no studies of the physical properties of the plasma producing 
the asymmetries. Here we identify regions of asymmetric profiles in the outflows of 
AR 10978 using an asymmetric Gaussian function and extract 
the intensities of the faint 
component using multiple Gaussian fits. We then derive the temperature structure
and chemical composition of the plasma producing the asymmetries. We find that the asymmetries are dependent on
temperature, and are 
clearer and stronger in coronal lines. 
The temperature distribution peaks around 1.4{\em--}1.8\,MK with an emission measure at least an order of magnitude
larger than that at 0.6MK.
The first ionization potential bias is found to be 3{\em--}5, implying that 
the high speed component of the outflows may also contribute to the slow speed wind. Observations and models
indicate that it takes time for plasma to evolve to a coronal composition, suggesting that the material
is trapped on closed loops before escaping, perhaps by interchange reconnection. The results, therefore, 
identify the 
plasma producing the asymmetries as having a {\it coronal origin}.
\end{abstract}
\keywords{Sun: corona---Sun: abundances---solar wind}

\section{Introduction}
Understanding how magnetic energy is generated, transported, and dissipated to 
heat the solar corona, where the solar wind originates and how it is 
accelerated, remain the most important unsolved problems in solar astrophysics. 
The {\it Hinode} mission \citep{kosugi_etal2007} 
was designed to investigate these topics and its instruments 
are providing many insights. The EUV Imaging Spectrometer 
\citep[EIS,][]{culhane_etal2007a} is able to measure the solar spectrum in the 
171{\em--}212\,\AA\, and 245{\em--}291\,\AA\, wavelength ranges and provides measurements that can be used to 
derive plasma properties such as density, temperature, emission measure (EM), and 
chemical composition. An important advance made possible by EIS is the ability 
to measure coronal line profiles at high spectral  resolution (22\,m\AA). These 
measurements have shown that the line profiles often have faint blue wing 
asymmetries \citep{hara_etal2008}.

The origin and importance of these asymmetries has not been clearly established. 
Nanoflare reconnection models predict that very hot lines should show weak blue 
wing enhancements
\citep{patsourakos&klimchuk_2006,klimchuk_2006}, while competing chromospheric 
jet models sugest that upflows should be seen at loop footpoints at 
`warm' temperatures \citep{aschwanden_etal2007b}. 
Active region outflows 
\citep{sakao_etal2007,harra_etal2008,delzanna_2008,doschek_etal2008} clearly
show asymmetries \citep{bryans_etal2010,doschek_2012}, so they could
have some connection to the solar wind. They are less easy to associate with 
specific features in other areas of an active region \citep{doschek_2012}, but several studies have 
indicated that they may be seen in and around plage regions and loop footpoints 
\citep{hara_etal2008,depontieu_etal2009,depontieu_etal2011,peter_2010}, though
they may be transient \citep{ugarteurra&warren_2011}. 
There are many ongoing investigations 
using EIS data 
\citep{mcintosh&depontieu_2009b,mcintosh&depontieu_2009a,tian_etal2011a,nishizuka&hara_2011,martinez_etal2011}.
One difficulty is that the asymmetries, even when present, account for less than $\sim$ 20\% of 
the total area of a line profile, so multiple Gaussian fitting or novel 
analysis techniques have been developed to extract as much information as 
possible from the profiles \citep{depontieu_etal2009,bryans_etal2010,peter_2010}.

\begin{figure*}
\centering
\includegraphics[width=0.45\linewidth,viewport= 12 45 355 310,clip]{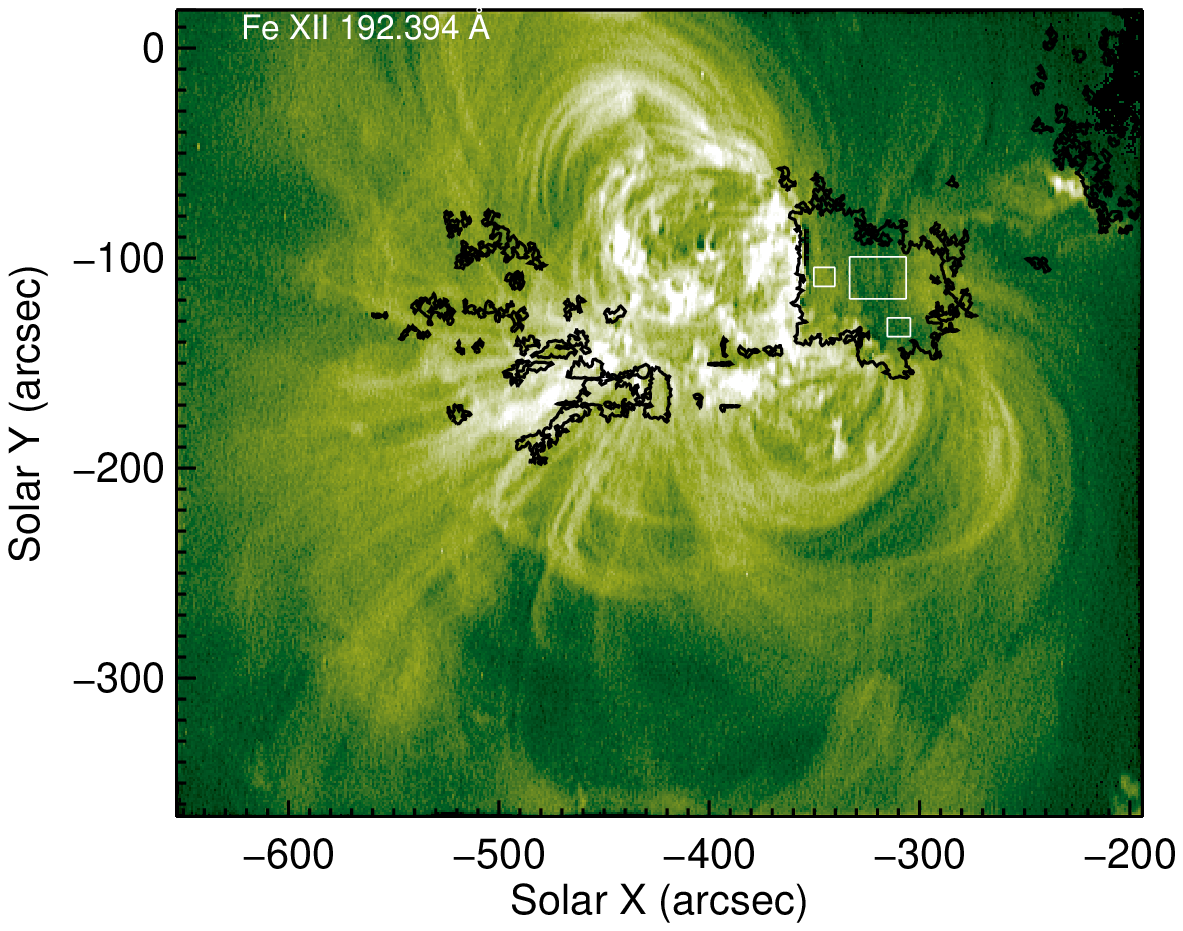}
\includegraphics[width=0.45\linewidth,viewport= 12 45 355 310,clip]{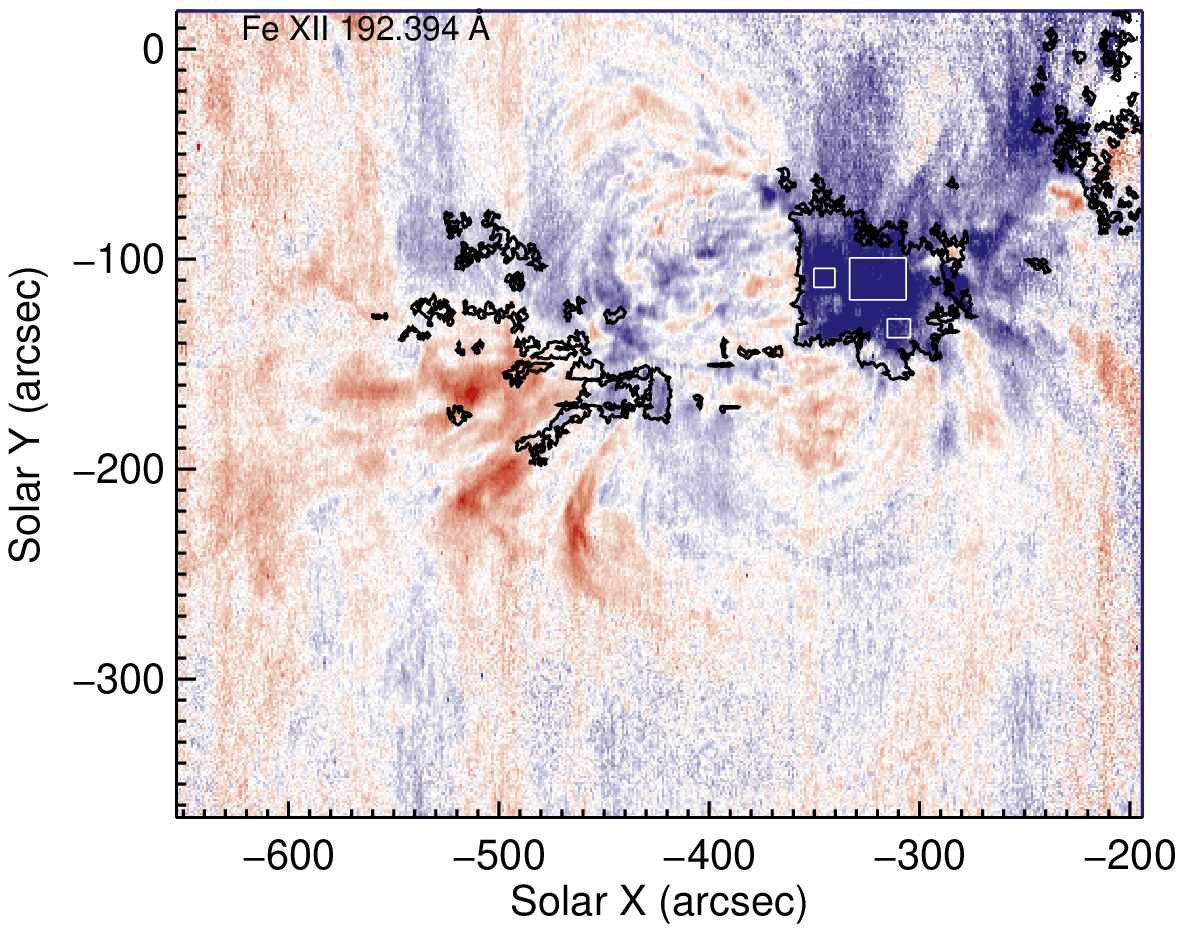}
\caption{EIS
\ion{Fe}{12} 192.394\,\AA\, image of AR 10978 (left) and Doppler velocity map (right).
The area used for analysis of the outflow is shown by the large white box. The size of
the box is 28$''$ $\times$ 21$''$. Two smaller boxes show other areas discussed later.
Contours indicating the locations where there is a $>$ 10\% blue-wing asymmetry
in the line profile are also shown. Contours shorter than 2\% of the largest 
one are omitted for clarity.  
The intensity image was treated with a Gaussian smoothing filter for 
presentation. 
\label{fig1}}
\end{figure*}

To date, studies of the asymmetries have focused on their location 
and velocity characteristics, but there have been no studies of the 
physical properties (temperature, chemical composition) of the plasma producing them. These diagnostics 
could provide new clues for coronal heating, solar wind, or 
chromosphere-corona connection studies, and this is the purpose of the 
investigation presented here. We focus on the outflow regions of AR 10978
that was observed by EIS in December 2007. AR 10978 was chosen to assess the importance  
of the asymmetric component since the outflows of this region have 
previously been linked to the solar wind through abundance measurements \citep{brooks&warren_2011}.
Furthermore, strong blue wing asymmetries are clearly seen
\citep{bryans_etal2010}, enabling a more robust analysis in what is a 
difficult measurement. We use 
an asymmetric Gaussian function to detect deviations from a symmetric profile and then extract the 
intensities of the asymmetric component for a range of lines from 
\ion{Si}{10}, \ion{S}{10}, and \ion{Fe}{8}{\em--}\ion{Fe}{17} using multiple component fitting. We use these 
measurements to derive temperatures, emission measures, and relative 
abundances for the asymmetric component.

\section{Data Reduction and Analysis}
We analyze observations of AR 10978 obtained on December 10{\em--}15, 2007. The data were
processed using standard software. Context intensity and Dopppler velocity images
are shown in Figure \ref{fig1}.  
A large area of outflow is seen on the Western side of the AR. 

To detect regions of non-Gaussian shaped line profiles, we fit the data with an asymmetric Gaussian
function of the form
\begin{equation}
I(\lambda) = b(\lambda) + \left\{
 \begin{array}{rl}
 A\exp\left(-\frac{(\lambda-\lambda_0)^2}{2\sigma_L^2}\right) &
\mbox{$\lambda\le\lambda_0$} \\
 A\exp\left(-\frac{(\lambda-\lambda_0)^2}{2\sigma_R^2}\right) &
\mbox{$\lambda>\lambda_0$  }
 \end{array}
 \right.
\end{equation}
where $I(\lambda)$ is the intensity at wavelength position $\lambda$, $\lambda_0$ is the centroid of the asymmetric Gaussian,
$A$ is the area under the curve, $\sigma_L$ and $\sigma_R$ are the widths of the Gaussian functions that 
reproduce the left and right wings of the profile, and $b(\lambda)$ is a polynomial representing the background
which is assumed to be linear. The asymmetry $\alpha$ is defined as $(\sigma_L-\sigma_R)/(\sigma_L+\sigma_R)$.
This function collapses to a Gaussian if there is no asymmetry. Contours where $\alpha >$ 0.1 (greater than 10\%
asymmetry) are overlaid on the intensity and velocity maps in Figure \ref{fig1}. They clearly show asymmetries
in the outflows. Only a small fraction of the pixels in this raster scan ($\sim$ 9\%) show line profile asymmetries
as large as this, and only $\sim$ 2\% show asymmetries greater than 20\%. It is clear that the asymmetries are 
not very prevalent, at least in this region. 

To increase the signal for fitting the weaker lines, we average the spectra over boxed regions within the contours.
An example box within the Western outflow on December 10 is shown in Figure \ref{fig1}.
To enable comparison of the results with our previous work on this region, we first performed
single Gaussian fits to the averaged spectra and 
extracted the intensities of a subset of 
\ion{Si}{10}, \ion{S}{10}, and \ion{Fe}{8}{\em--}\ion{Fe}{17} lines. 

\begin{figure*}
\centering
\includegraphics[width=0.32\linewidth,clip]{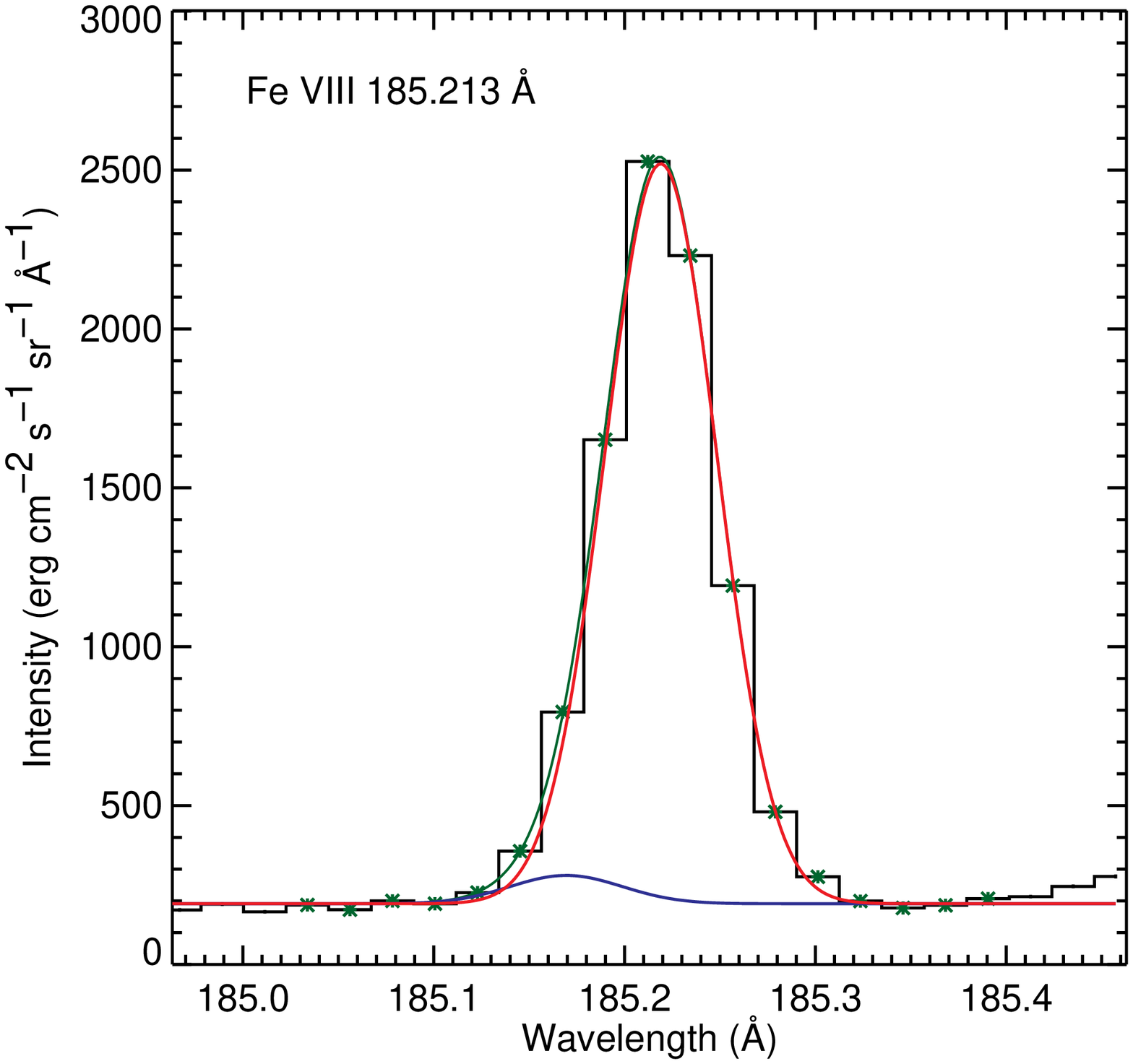}
\includegraphics[width=0.32\linewidth,clip]{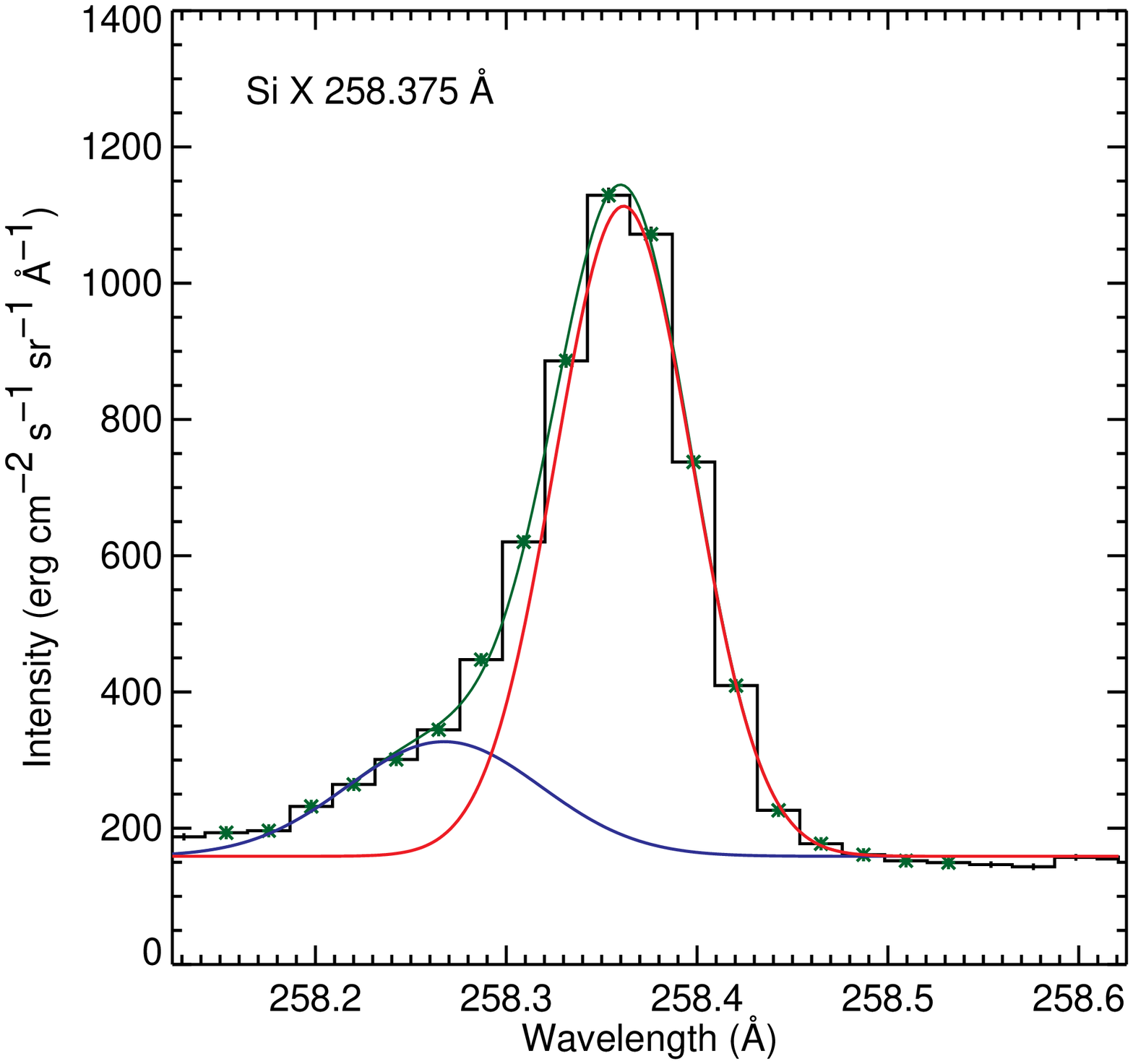}
\includegraphics[width=0.32\linewidth,clip]{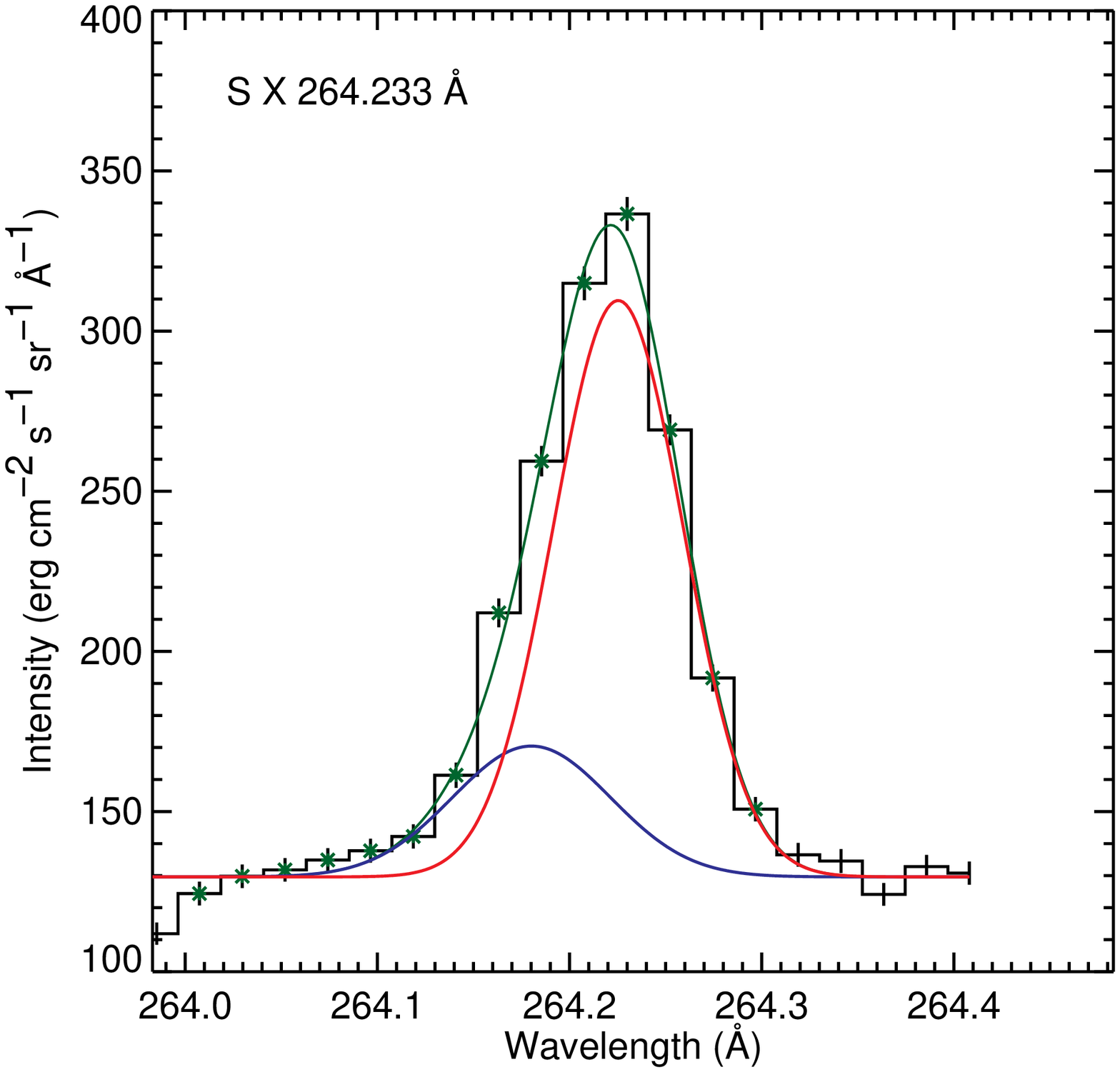}
\includegraphics[width=0.32\linewidth,clip]{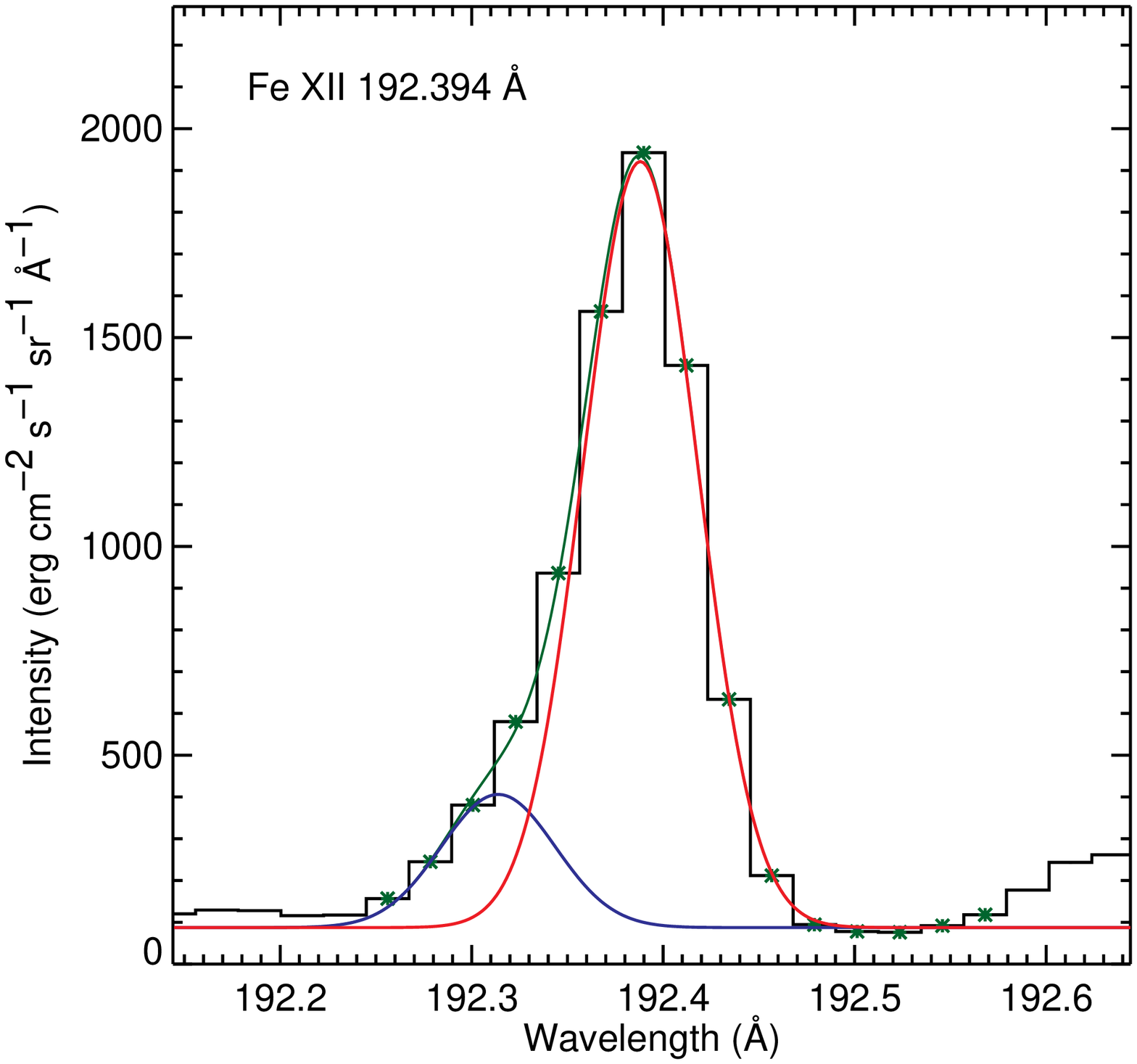}
\includegraphics[width=0.32\linewidth,clip]{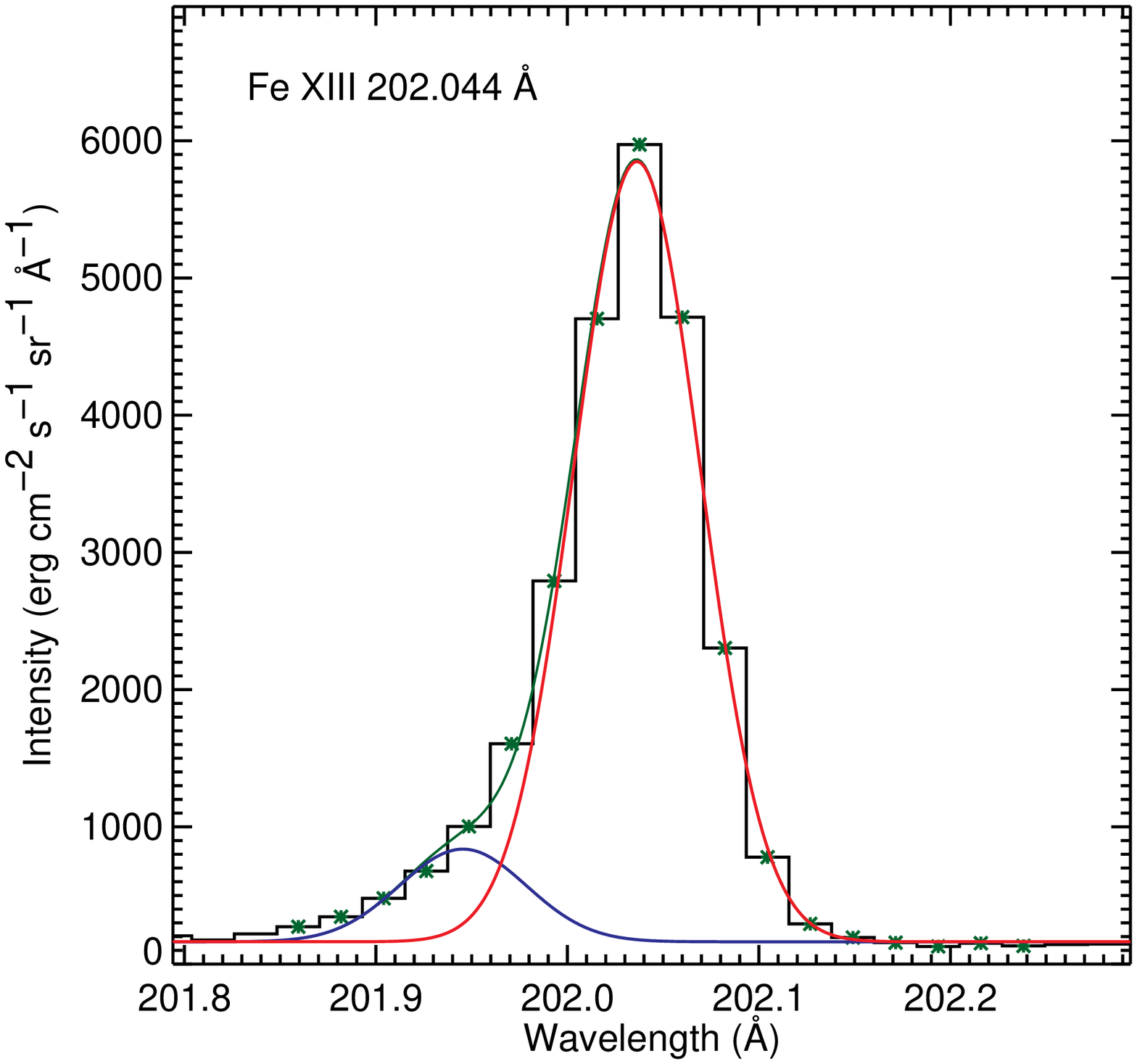}
\includegraphics[width=0.32\linewidth,clip]{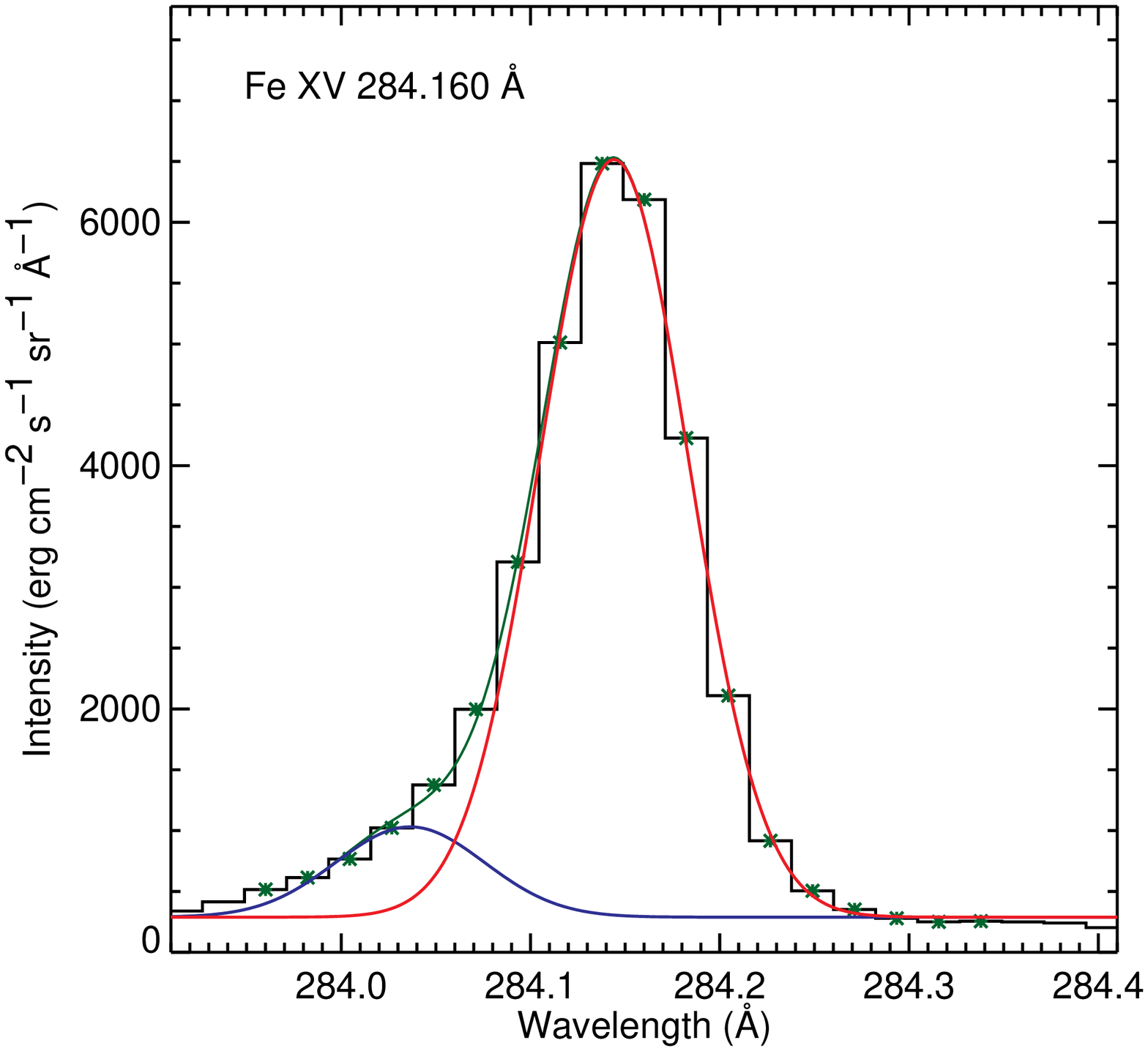}
\caption{Multiple component fits to the spectral line profiles for the boxed region of the
Eastern outflow shown in Figure \ref{fig1}. Fits to \ion{Fe}{8} 185.213\,\AA, \ion{Si}{10} 258.375\,\AA,
\ion{S}{10} 264.233\,\AA, \ion{Fe}{12} 192.394\,\AA, \ion{Fe}{13} 202.044\,\AA, and \ion{Fe}{15} 284.160\,\AA\, are shown, 
spanning a range of temperatures
from 0.4--2.2MK. 
The histograms and green stars show the observed spectra and the solid green lines show the composite
fit to the data. The red line shows the primary component of the fit, and the blue line shows the 
minor component (blue wing asymmetry).  
\label{fig2}}
\end{figure*}

To extract the intensities of the asymmetric components we performed multiple Gaussian fits. 
\citet{peter_2010} pointed out that fitting multiple Gaussians implicitly assumes that the emission
is the combined contribution of a fixed number of spatial components. This may not be true of the 
real emission, as also noted by \citet{bryans_etal2010}, which could be produced by a continuous distribution
of plasma velocities. It is important, therefore, to consider uncertainties in the fitting procedure we adopt.
We have found that fitting two components of equal width leads to consistent
and reasonable results in general. 
\citet{tian_etal2011b} have also found that the
minor component usually has a Gaussian width comparable to the main component. There are circumstances, however, 
where this assumption is unsatisfactory: 
if the asymmetric component has a weak high speed tail, the restriction of fixed widths can lead to an 
underestimation of the minor component intensity. This sometimes 
affects the critical \ion{Si}{10} 258.375\,\AA\,
line but not \ion{S}{10} 264.233\,\AA. Since \ion{S}{10} 264.233\,\AA\, is weaker, the high
speed tail may be lost in the noise on these occasions,
and we prefer to make a direct comparison by applying the same method to both lines.
When \ion{S}{10} 264.233\,\AA\, is strong enough, however, we assume that
any high speed tail should be visible, and we allow the width of the minor component to vary from the major
component. In summary, in most cases we fit two components of fixed width to all the lines. In a few
cases (3), however, we allow the minor component width of the \ion{Si}{10} 258.375\,\AA\, and \ion{S}{10} 264.233\,\AA\, lines
to vary. Even in these cases all the \ion{Fe}{0} lines are fitted with fixed widths. 
Examples for spectral lines covering 
a range of temperatures are shown in Figure \ref{fig2}. 

Once the intensities are extracted we perform the EM calculations. Recent analysis has shown that
the outflows are more complex than previously thought. \citet{warren_etal2011a} and \citet{young_etal2012} have found that lower temperature
lines show downflows on loop structures that are unassociated with the upflows. To account for this,
we only use lines where the total outflow emission is blue-shifted. If a lower temperature line is red-shifted,
the intensity is set to zero and a 3$\sigma$ error is used for the uncertainty, where $\sigma$ is the error in the
measured background intensity from the fit.
This correction is also applied to any line that is too weak to obtain a reliable
fit. All the other lines have the instrument calibration uncertainty \citep{lang_etal2006} added in quadrature 
to the intensity measurement error. 

To derive the EM we use the Markov-Chain
Monte Carlo (MCMC) algorithm distributed with the PINTofALE spectroscopy package \citep{kashyap&drake_1998,kashyap&drake_2000}
and compute contribution functions from the CHIANTI v7.0 database \citep{dere_etal1997,landi_etal2012}. We adopt
the CHIANTI ionization fractions and the photospheric abundances of \citet{grevesse_etal2007}. The contribution
functions are calculated at the electron density derived using the reliable \ion{Fe}{13} 202.044\,\AA/203.826\,\AA\, ratio
for the total emission. The \ion{Fe}{13} 203.826\,\AA\, line is blended on the blue side, however, with \ion{Fe}{12} 203.734\,\AA,
making it difficult to extract any asymmetric component with reasonable uncertainties. Therefore, we use the same 
density to calculate the contribution function for the asymmetric component. We explored the possibility of using
the \ion{Fe}{12} 186.880\,\AA/192.394\,\AA\, ratio instead and were able to extract the asymmetric component
for the 186.880\,\AA\, line, but the known discrepancy between densities derived using \ion{Fe}{12} and \ion{Fe}{13}
\citep{young_etal2009} also became apparent for these outflows and introduced uncertainties around the temperature of formation
of the critical \ion{S}{10} and \ion{Si}{10} lines used for the abundance measurements. For direct comparison between the asymmetric component and the
bulk outflow, and also for benchmarking against our previous work \citep{brooks&warren_2011}, we decided to only
use \ion{Fe}{13}. 

\begin{figure*}
\centering
\includegraphics[width=0.48\linewidth]{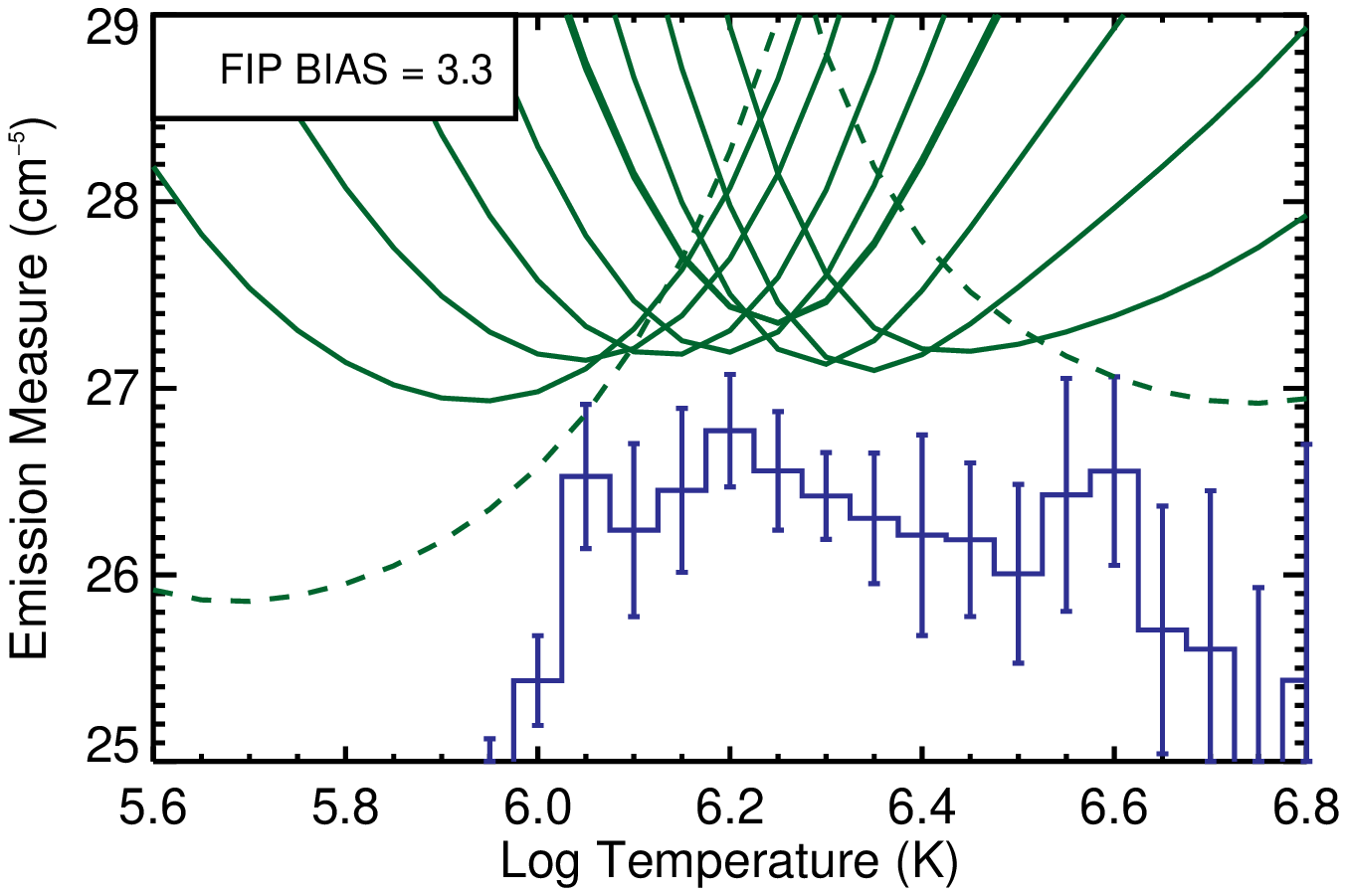}
\includegraphics[width=0.48\linewidth]{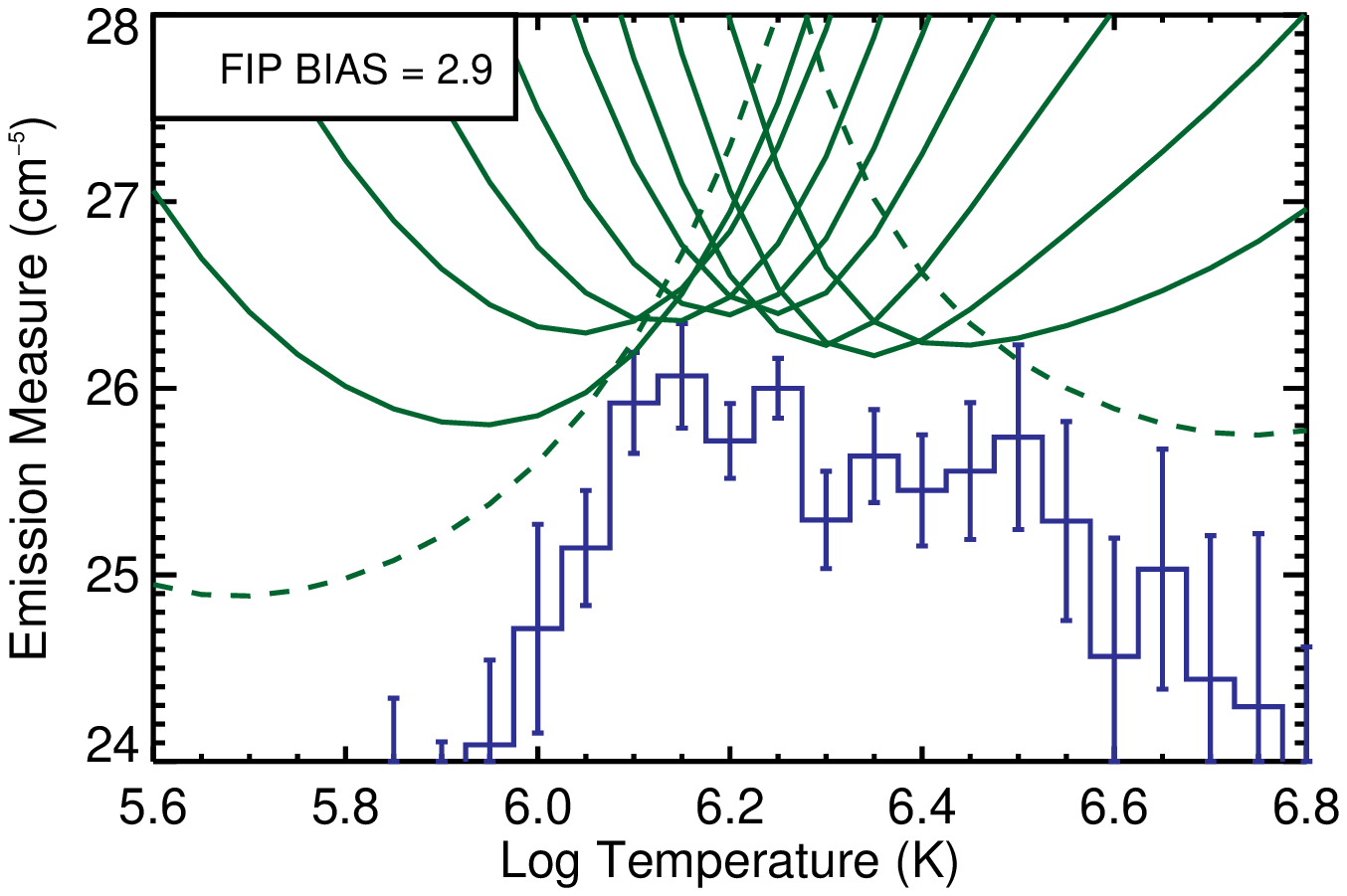}
\caption{Emission measure results for all the emission in the outflow (left) and for the asymmetric component only (right).
The blue line shows the best fit solution. 1$\sigma$ error bars computed from 100 Monte Carlo simulations using
perturbed intensities are indicated. These show the possible spread in results. 
The green lines are emission measure loci curves indicating where the 
\ion{Fe}{0} lines constrain the solution. The dashed curves denote lines where the computed intensity was used (because the observed intensity 
was set to zero; see text). 
\label{fig3}}
\end{figure*}

\begin{deluxetable*}{lccccccccccc}
\tabletypesize{\scriptsize}
\tablewidth{0pt}
\tablecaption{Differential Emission Measure Model}
\tablehead{
\multicolumn{1}{l}{} & 
\multicolumn{4}{c}{Total Outflow Emission} & 
\multicolumn{1}{c}{} & 
\multicolumn{4}{c}{Asymmetric Component} \\ 
[.3ex]\cline{2-5}\cline{7-12} \\[-1.6ex] 
\multicolumn{1}{l}{Line ID} &
\multicolumn{1}{c}{I$_{obs}$} &
\multicolumn{1}{c}{$\sigma_I$} &
\multicolumn{1}{c}{I$_{dem}$} &
\multicolumn{1}{c}{R(\%)}  &
\multicolumn{1}{c}{ }  &
\multicolumn{1}{c}{I$_{obs}$} &
\multicolumn{1}{c}{$\sigma_I$} &
\multicolumn{1}{c}{I$_{dem}$} &
\multicolumn{1}{c}{R(\%)} &
\multicolumn{1}{c}{ }  &
\multicolumn{1}{c}{P(\%)} \\
}
\startdata
Fe VIII 185.213 &   0.0 &   5.3 &  10.8 &  0.0 & &   0.0 &   5.9 &   1.2 &  0.0 & &  0.0 \\
  Fe IX 197.862 &  32.1 &   7.1 &  21.1 & 34.3 & &   2.4 &   0.6 &   2.6 &  6.9 & &  7.4 \\
   Fe X 184.536 & 161.4 &  35.5 & 131.6 & 18.5 & &  22.6 &   5.6 &  18.2 & 19.7 & & 14.0 \\
  Fe XI 188.216 & 278.7 &  61.3 & 313.4 & 12.5 & &  42.2 &   9.3 &  45.9 &  8.9 & & 15.1 \\
 Fe XII 192.394 & 150.4 &  33.1 & 177.7 & 18.1 & &  23.8 &   5.3 &  24.4 &  2.2 & & 15.8 \\
Fe XIII 202.044 & 496.6 & 109.3 & 378.7 & 23.8 & &  56.0 &  12.4 &  46.8 & 16.5 & & 11.3 \\
Fe XIII 203.826 & 195.1 &  43.0 & 148.7 & 23.8 & & -- & -- & -- & -- & & -- \\
 Fe XIV 264.787 &  99.0 &  21.8 & 107.0 &  8.1 & &  12.5 &   2.8 &  12.6 &  0.7 & & 12.6 \\
  Fe XV 284.160 & 609.4 & 134.1 & 655.1 &  7.5 & &  73.1 &  16.1 &  76.7 &  5.0 & & 12.0 \\
 Fe XVI 262.984 &  25.0 &   5.5 &  24.6 &  1.7 & &   2.7 &   0.8 &   2.6 &  3.6 & & 10.8 \\
 Fe XVII 254.87 &   0.0 &   7.5 &   1.2 &  0.0 & &   0.0 &   8.9 &   0.1 &  0.0 & &  0.0 \\
    S X 264.233 &  18.5 &   0.6 & -- & -- & &   4.2 &  17.3 & -- & -- & & 22.8 \\
   Si X 258.375 &  99.6 &   1.1 & -- & -- & &  21.5 &   0.9 & -- & -- & & 21.6 \\
\enddata
\tablenotetext{}{I$_{obs}$ is the observed intensity. $\sigma_I$ is the intensity error with the
photometric calibration uncertainty added in quadrature. I$_{dem}$ is the calculated intensity.
R is the absolute difference between the observed and calculated intensity. P is the ratio of the 
intensity of the asymmetric component to that of the total outflow emission. R and P are expressed as percentages.}
\label{tab1}
\end{deluxetable*}

The methodology used to derive the first ionization potential (FIP) bias
has been presented by \citet{brooks&warren_2011}, and we have found our chosen lines to be 
reliable in separate emission measure analyses of a variety of solar regions
\citep{warren&brooks_2009,brooks_etal2009,brooks_etal2011,warren_etal2011b}. Many
of them have also been independently verified in other studies \citep{testa_etal2011,winebarger_etal2011,tripathi_etal2011}, strengthening our confidence in the atomic data. Furthermore, in \citet{brooks&warren_2011} we tested the method
by applying the analysis to eight polar coronal hole observations and verified that photospheric abundances were
obtained, as expected for the presumed source of the fast solar wind.
Briefly, the \ion{S}{10} 264.233\,\AA\, and \ion{Si}{10} 258.375 lines
are close in both wavelength and formation temperature, so the ratio
is a potentially useful diagnostic for measuring abundances. Unfortunately, the ratio of these lines has some temperature and density
sensitivity. To account for the temperature sensitivity we convolve the ratio with the EM distribution derived for the outflows. 
The EM is reconstructed from the \ion{Fe}{0} 
lines only. This minimizes uncertainties due to the choice of elemental abundances. The EM is then scaled to match
the \ion{Si}{10} 258.375\,\AA\, intensity. This scaling is introduced to account for uncertainty in the \ion{Fe}{0}/\ion{Si}{0}
abundance ratio and is generally small (less than 40\% in the majority of cases). The intensity of the 
\ion{S}{10} 264.233\,\AA\, line is then predicted. Since \ion{S}{0} is a high FIP element and \ion{Si}{0} is a low FIP 
element, the ratio of predicted to observed \ion{S}{10} 264.233\,\AA\, intensity is the FIP bias. 
The inferred FIP bias
is not very sensitive to the assumed density. Changing the density by a factor of 3 changes the derived FIP bias by
approximately 40\%.

\section{Results}

It is evident from Figure \ref{fig2} that the 
magnitude of the asymmetric component is strongly dependent on 
temperature. 
Compare, for example, the fits to \ion{Fe}{8} 185.213\,\AA\, and \ion{Fe}{15} 284.160\,\AA.
Table \ref{tab1} shows both the total intensity and that of the asymmetric component for all the lines
used for the EM analysis of the region in Figure \ref{fig1}. For this example, the asymmetric 
component amounts to 7\% of the total emission at 0.7\,MK, increasing to 16\% at 1.6\,MK. This result
is broadly representative of all the regions we have studied.

\begin{deluxetable*}{lcccccccc}
\tabletypesize{\scriptsize}
\tablewidth{0pt}
\tablecaption{Outflow Properties}
\tablehead{
\multicolumn{1}{l}{} & 
\multicolumn{4}{c}{Total Outflow Emission} & 
\multicolumn{4}{c}{Asymmetric Component} \\ 
[.3ex]\cline{2-5}\cline{6-9} \\[-1.6ex] 
\cline{1-9} \\[-1.6ex]
\multicolumn{1}{l}{Data} &
\multicolumn{1}{c}{Outflow} &
\multicolumn{1}{c}{T$_p$} &
\multicolumn{1}{c}{EM$_p$} &
\multicolumn{1}{c}{FIP}  &
\multicolumn{1}{c}{T$_p$} &
\multicolumn{1}{c}{EM$_p$} &
\multicolumn{1}{c}{FIP} &
\multicolumn{1}{c}{$\Delta\log$EM} \\
}
\startdata
eis\_l1\_20071210\_0019 & West & 1.6 & 27.6 & 3.3 & 1.4 & 26.7 & 2.9 & 3.4 \\
eis\_l1\_20071210\_0019 & West & 1.8 & 28.1 & 3.5 & 1.4 & 26.9 & 3.3 & 2.5 \\
eis\_l1\_20071210\_0019 & West & 1.4 & 27.6 & 3.5 & 1.8 & 26.6 & 3.4 & 2.8 \\
eis\_l1\_20071211\_1025 & West & 3.2 & 27.8 & 3.3 & 1.6 & 26.8 & 3.0 & 2.7 \\
eis\_l1\_20071211\_1025 & East & 3.2 & 28.3 & 4.1 & 1.6 & 27.1 & 4.7 & 4.3 \\
eis\_l1\_20071211\_1025 & West & 1.8 & 28.1 & 3.2 & 2.0 & 27.0 & 3.3 & 2.8 \\
eis\_l1\_20071211\_1025 & West & 1.8 & 27.5 & 3.3 & 1.4 & 26.6 & 2.8 & 2.1 \\
eis\_l1\_20071211\_1025 & West & 1.8 & 27.8 & 3.3 & 1.6 & 26.6 & 3.8 & 3.2 \\
eis\_l1\_20071212\_1143 & West & 1.6 & 27.6 & 2.9 & 1.6 & 26.7 & 3.0 & 3.2 \\
eis\_l1\_20071212\_1143 & East & 1.6 & 28.0 & 4.6 & 1.4 & 26.9 & 4.9 & 3.7 \\
eis\_l1\_20071212\_1143 & West & 1.8 & 28.0 & 3.6 & 1.6 & 26.8 & 3.1 & 2.4 \\
eis\_l1\_20071212\_1143 & East & 2.2 & 27.9 & 3.4 & 1.8 & 26.7 & 3.0 & 4.1 \\
eis\_l1\_20071213\_1218 & West & 3.2 & 28.2 & 2.9 & 1.6 & 26.7 & 4.9 & 4.1 \\
eis\_l1\_20071213\_1218 & East & 1.6 & 27.7 & 4.0 & 1.4 & 26.8 & 3.8 & 2.1 \\
eis\_l1\_20071215\_0013 & East & 1.8 & 27.8 & 3.1 & 1.4 & 26.9 & 4.2 & 3.2 \\
\enddata
\tablenotetext{}{The EIS datasets used are given along with the location of the outflow.
The peak temperature (in MK), peak emission measure (cm$^{-3}$), and FIP bias are given for
calculations using both the total emission and the asymmetric component only. The ratio of 
the peak emission measure to that measured at 0.6MK is also shown.}
\label{tab2}
\end{deluxetable*}

The derived EM distributions are shown in Figure \ref{fig3}.
The left panel shows the EM for the 
whole outflow at a derived density of $\log$ (n$_e$/cm$^{-3}$) = 8.55, and
the right panel shows the EM for the plasma producing the asymmetries.
Table \ref{tab1} shows that the
observed intensities are reproduced by the EM model to within 30\% for all lines except 
\ion{Fe}{9} 197.862\,\AA\, which is a little too bright in the bulk emission case. 
This is seen in another example and in fact more often in 
\ion{Fe}{10} 184.536\,\AA. 
As discussed above, the transition
from downflows on closed structures to outflows on open field lines takes place in this temperature range, and 
our method of abruptly including or excluding lines based on Doppler shift probably does not account 
perfectly for the gradual transition from emission dominated by one type of structure to the next. 
In fact it is difficult to select boxes on the Eastern side of the AR that are both within the outflow
and asymmetry regions and are clear of the fans. This is the reason that we have selected relatively
more areas from the Western side in the following analysis of a sample of outflows.
Note, however, that 
there is an \ion{Fe}{11} line at 184.412\,\AA\, that is easily separable from \ion{Fe}{10} 184.536\,\AA\, in
normal solar conditions, but becomes blended with the profile when there is a high speed component. This could affect
the \ion{Fe}{10} intensity measurements. Also, there is some uncertainty as to 
whether the \ion{Fe}{8} and \ion{Fe}{9} intensities can simultaneously be reproduced by the EM of every type of solar
feature \citep{schmelz_etal2012}. 

To study the generality of the results, we repeated our analysis for 14 additional regions in these AR
observations. The chosen regions range in size from 10$''$ $\times$ 8$''$ to 28$''$ $\times$ 16$''$. Our results, which are summarized in Table \ref{tab2}, show that 
the temperature distributions for the outflows peak in the range 1.4{\em--}3.2\,MK
with an emission measure of $\log$ (EM/cm$^{-5}$) = 27.5{\em--}28.3. The measured densities (not shown) are $\log$ (n$_e$/cm$^{-3}$) = 8.4{\em--}8.9. 
These numbers are consistent with 
previous EM analysis of AR outflows \citep{brooks&warren_2011,slemzin_etal2012}. 
The derived FIP bias falls in the 
range 2.9{\em--}4.6. 

The EM distributions for the asymmetric component appear to peak in a narrower temperature 
range: 1.4{\em--}2.0\,MK, and 
they show rapidly decreasing emission measure at temperatures
below 1\,MK. We characterize this fall off by measuring the ratio of the EM
at the peak to that at 0.6\,MK. In all cases 
the EM is at least two orders of magnitude larger at the peak temperature.
The derived FIP bias for the asymmetric component falls in the range 2.8{\em--}4.9, which is similar
to that of the bulk outflow. 

Although we excluded lines with a bulk red-shifted profile, they often have weak blue wing emission,
see e.g., \ion{Fe}{8} 185.213\,\AA\, in Figure \ref{fig2}. We therefore re-computed the results for the
asymmetric component including these lines. We found that they are far enough 
away in temperature that the peak emission measure, temperature, and FIP bias are not affected significantly
if they are included. In contrast, $\Delta \log EM$ obviously changes significantly for individual cases, though,
as discussed, it is unclear if the low temperature EM becomes too large when contaminated by emission from other
structures. The range of results remains broadly the same, however, and in all cases the EM is still at least an order of
magnitude larger at the peak temperature. 

\section{Summary and discussion}
We have measured the physical properties of several outflow areas in AR 10978 during its disk passage
in December 2007. 
We also extracted measurements
for the plasma producing the faint blue wing asymmetries in the EIS line profiles.
Results for the bulk outflow are consistent with those of previous studies. 

The new measurements for the asymmetric component show that the magnitudes of the 
asymmetries are strongly dependent on temperature and peak in the corona. 
The emission of the source plasma appears also to be dominated by, and come from, a coronal
temperature range. The relative abundance measurements indicate 
a similar 
FIP fractionation value to that of the bulk outflow.

These results have two important consequences. First,
the FIP bias values (3{\em--}5) are slightly higher than but broadly similar to those of the slow speed solar wind \citep[3{\em--}4,][]{vonsteiger_etal2000}. 
The higher value probably reflects the difficulty of making this measurement, but 
indicates that the high speed component of the outflows could also be a contributor
and source of the slow speed wind.

Second, in relation
to coronal heating studies, 
previous work has indicated that even if rapid fractionation occurs 
locally, it takes at least several hours for the plasma to evolve to a coronal composition on a large
scale \citep{feldman&widing_2003}. Other work has suggested that the material is stored on coronal loops and is
later released by interchange reconnection with open field lines \citep{wang_etal1996,schwadron_etal1999}. 
Our FIP bias results therefore imply that
the asymmetries in the AR outflow are likely to be 
produced by a mechanism that releases plasma of {\it coronal} origin. 
This is also consistent with our determination that the EM distributions are dominated by coronal
emission.

Finally, it is unclear whether our results could support a picture whereby the asymmetries are a signature of plasma that is rapidly
injected into the corona from the chromosphere. In addition to the apparent coronal source,
recent models suggest that fractionation is harder and therefore reduced for
plasma that passes rapidly through the fractionation region \citep{laming_2004,laming_2012}.
In the most obvious case, we would expect to measure a photospheric
composition, and therefore the  
\ion{S}{0} line would be 
relatively more asymmetric than the \ion{Si}{0} line; the opposite of what we see. 
We stress, however, that this initial study has been of asymmetric line profiles averaged
over relatively large areas in AR outflows only. 
Similar studies on smaller scales may also be possible in bright moss or plage, for example,
since good line profiles are obtained with EIS even with short exposures \citep{brooks&warren_2009}. 
Linking these weak asymmetries to small scale chromospheric jets may require a higher sensitivity
spectrometer with better spatial resolution covering the whole atmosphere from chromosphere to
corona. Such a spectrometer is proposed for Solar-C \citep{teriaca_etal2011}.

\acknowledgements
We thank Martin Laming for helpful discussions.
This work was performed under contract with the Naval Research Laboratory and
was funded by the NASA {\it Hinode} program.
{\it Hinode} is a Japanese mission developed and launched by ISAS/JAXA,
with NAOJ as domestic partner and NASA and STFC (UK) as international partners.
It is operated by these agencies in co-operation with ESA and NSC (Norway).

\end{document}